# *Bistability in mushroom-type metamaterials*

David E. Fernandes[a], and Mário G. Silveirinha[a,b] [1]

[a]*University of Coimbra, Department of Electrical Engineering and Instituto de Telecomunicações, 3030-290, Coimbra, Portugal*

[b]*University of Lisbon, Instituto Superior Técnico - Avenida Rovisco Pais, 1, 1049-001 Lisboa, Portugal*

**Abstract**

Here, we study the electromagnetic response of asymmetric mushroom-type metamaterials loaded with nonlinear elements. It is shown that near a Fano resonance these structures may have a strong tunable, bistable, and switchable response and enable giant nonlinear effects. Using an effective medium theory and full wave simulations, it is proven that the nonlinear elements may allow the reflection and transmission coefficients to follow hysteresis loops, and to switch the metamaterial between "go" and "no-go" states similar to an ideal electromagnetic switch.

---

[1] To whom correspondence should be addressed: E-mail: mario.silveirinha@co.it.pt

# I. Introduction

The uniaxial wire medium consists of a set of infinitely long parallel metallic wires embedded in a dielectric host. This metamaterial has attracted a lot of attention due to its numerous applications in the microwave, THz and optical frequency bands [1-25]. Notably, the wire medium is characterized by a strongly nonlocal (spatially dispersive) response, even in the quasi-static limit [1, 5]. It was shown in Ref. [10] that the nonlocal response can be tamed by loading the wires with metallic patches. This idea was further developed in subsequent works [5, 13-15]. These structures are usually known as mushroom-type metamaterials. Moreover, it was also shown that the electromagnetic response of the metamaterial can be tailored by loading the wires with lumped loads [16, 26, 27] or alternatively by misaligning the geometrical center of the patch elements with respect to the wires so that the structure becomes asymmetric [28, 29].

The research of nonlinear effects in wire media has been mainly focused on the propagation of light in structures formed by metallic nanowires embedded in a nonlinear Kerr-type dielectric host [30-33]. Moreover, in a recent series of works it was shown that mushroom-type ground planes with a nonlinear response can be used to absorb high-power signals [34-36]. In general, nonlinear metamaterials can enable bistable and multi-stable regimes, provide tunable and switchable responses, dramatically boost the sensitivity to nonlinear elements, and allow for frequency conversion and parametric amplification [37-52]. Furthermore, a nonlinear response may enable new functionalities such as "electromagnetic diodes" [53] and all-optical memories [54]. Recent surveys on the topic can be found in Refs. [55, 56].

Motivated by the many opportunities that are created by a nonlinear response, here we theoretically investigate the interaction between electromagnetic waves and asymmetric mushroom-type nonlinear metamaterials at microwave frequencies. Typically, a nonlinear microwave response is realized using nonlinear lumped elements, for instance a variable capacitance diode, commonly known as varactor. For simplicity of modeling, in our study we consider that the wires-to-patch connection is done through a small gap filled with a Kerr-type dielectric, which behaves as a nonlinear parallel-plate capacitor. To have pronounced nonlinear effects for normal plane wave incidence, the wire grid is misaligned with respect to the geometrical center of the metallic patches. We develop an effective medium framework to characterize the nonlinear response of the system in a stationary regime, and present a detailed numerical study and full wave simulations of the impact of varying the intensity of the excitation field. It is shown that the dynamic response of the mushroom metamaterial is bistable such that the structure may behave as an electromagnetic switch with "go" and "no-go" states. Moreover, near a Fano resonance there is a strong enhancement of the electric field that acts on the nonlinear element, and this property and the high sensitivity of the Fano resonance to the reactance of the lumped element effectively boost the nonlinear effects.

This paper is organized as follows. In Sec. II we introduce the system under study and highlight how a lumped load can enable controlling the response of the metamaterial. In Sec. III we study the nonlinear dynamics of the system. Using a homogenization model developed in an earlier work for the linear problem [29], we investigate the conditions required for the emergence of a bistable response and determine the hysteresis loops followed by the transmission and reflection coefficients. In Sec. IV we present full wave simulations of the dynamical response of the metamaterial, showing that by changing the intensity of an incoming wave it is possible to switch the system between different states and control the transmission level. In Sec. V the conclusions are drawn. In a time harmonic regime, it is assumed that the time variation is of the type $e^{-i\omega t}$, $\omega$ being the oscillation frequency.

## II. The Nonlinear Metamaterial Model

We consider a two-sided mushroom structure formed by a wire medium slab with thickness $h$ terminated with metallic patches, as depicted in Fig. 1a). The wires are arranged in a two-dimensional square lattice of period $a$ and are attached to the metallic patches either through a discrete lumped load $Z_L$ (bottom interface) or through a direct (ideal short-circuit) connection (top interface). It is also assumed that the connection point between the metallic wires and the patches is shifted with respect to the geometrical center of the patches, along the $x$-direction, as shown in Fig. 1b). It was demonstrated in previous works that when the patch grid and wire array are misaligned, the response of the mushroom is altered and new resonances appear, allowing the wire medium to strongly interact with an incident wave for *normal incidence* [28, 29]. An effective medium framework was put forward in Ref. [29] to characterize the scattering of plane waves by this two-sided mushroom structure in the linear regime. For the convenience of the reader, we present in the Appendix A a brief overview of the theory.

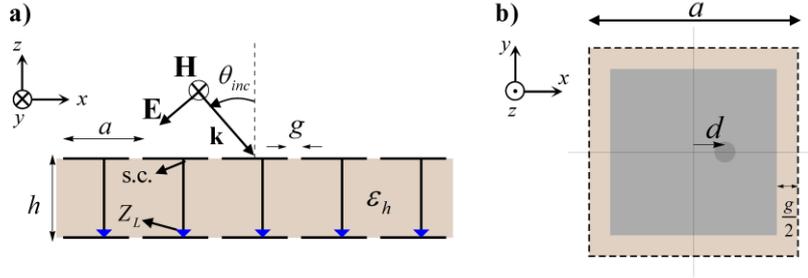

**Fig. 1** Geometry of the problem under study (two-sided mushroom slab) **a)** Side view: The wires are arranged in a square lattice with period $a$, embedded in a dielectric host with permittivity $\varepsilon_h$ and thickness $h$. The wires are connected to the bottom patch grid through lumped loads (blue arrows in the figure) and to the top patch grid through a short circuit (s.c.). The separation between adjacent patches is $g$. The slab is illuminated by a plane wave with an incidence angle $\theta_{inc}$. **b)** Top view of one cell of the two-sided mushroom structure: the wire is displaced by a distance $d$ with respect to the patch center along the positive $x$-direction.

The effective load impedance, $Z_{L,ef}$, depends on the lumped element impedance $Z_L$ and also on a parasitic capacitance and a parasitic inductance ($C_{par}$ and $L_{par}$), which are determined by the specific geometry of the connection point [16]. The effective impedance may be written as:

$$Z_{L,ef} = -i\omega L_{par} + \frac{1}{-i\omega C_{par} + 1/Z_L}, \qquad (1)$$

Here, we consider that the wire-to-patch connection is done through a nonlinear load whose impedance varies with the voltage drop $V_L$ at its terminals, $Z_L = Z_L(V_L)$.

It is well known [16, 26, 27, 29] that a lumped element may strongly affect the transmission and reflection coefficients of the metamaterial structure, even in the linear regime. To illustrate this, first we determine the scattering from a two-sided mushroom slab with thickness $h = 4a$, with $a = 1\,\text{cm}$ the period of the structure, for different inductive and capacitive loads. For simplicity, the wires are modeled as perfect electric conductors (PEC) and it is assumed that they are surrounded by air. The wire radius is $r_w = 0.025a$ and the offset distance between the wire and patch arrays is $d = a/4$. The gap between adjacent metallic plates is $g = 0.05a$. These structural parameters are used throughout the article. The metamaterial slab is excited with a plane wave propagating along the direction normal to the interface ($\theta_{inc} = 0°$). Figure 2 shows a comparison between the results obtained with the homogenization model and the commercial full-wave electromagnetic simulator CST Microwave Studio [57] for different values of load impedance. The values of the parasitic inductance and impedance are estimated using a curve fitting of the analytical results and the full-wave simulations (in the full wave simulations the load is placed in a tiny gap of $h/40$). In our case, the parasitic effects are modeled by $L_{par} = 77.5\,\text{pH}$ and $C_{par} = 40.9\,\text{fF}$.

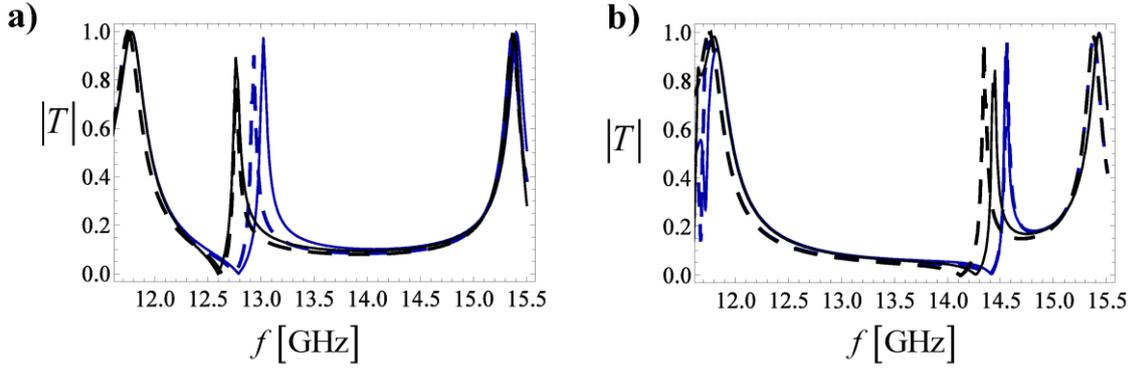

**Fig. 2** Amplitude of the transmission coefficient for the two-sided mushroom structure when the wires are connected to the patch grid through: **a)** inductive loads $L=1\text{nH}$ (blue curve) and $L=1.5\text{nH}$ (black curve) and **b)** capacitive loads $C=0.05\text{pF}$ (blue curve) and $C=0.075\text{pF}$ (black curve). Solid lines: homogenization model; Dashed lines: full-wave simulations.

The results reveal a good agreement between the effective medium and full-wave simulations, supporting the validity of the theory. The discrepancies between the two models are a consequence of the approximations implicit in the analytical model, which regards the metamaterial as an effective medium whereas the numerical simulations take into account all microscopic details of the structure. The observed discrepancies are somewhat similar to what is found in previous works [16, 26, 27]. It is seen in Fig. 2a that when the inductance of the lumped load increases, the second transmission resonance near 13GHz moves to lower frequencies. Similarly, larger capacitive loads shift the same resonance (now near 14.5 GHz) to lower frequencies (Fig. 2b). Therefore, the lumped load provides an interesting opportunity to control the scattering by the metamaterial slab. The second transmission resonance has a Fano-type lineshape [58, 59], and hence the transmission coefficient may be rather sensitive to a variation of the load reactance near this resonance. Fano resonances may emerge in wire media when a narrow quadrupole-type resonance interferes with a broad dipole-type resonance [60, 61].

In this work we aim to take advantage of the response of a nonlinear load. In our model, it is supposed that the lumped load consists of a parallel plate-type capacitor filled with a nonlinear Kerr-type dielectric (Fig. 3). Thus, the impedance of the lumped element may be estimated using:

$$Z_L = \frac{1}{-i\omega C_L} \quad (2)$$

where $C_L = \varepsilon_0 \varepsilon_{cap} \frac{A_{cap}}{t_{cap}}$ is the capacitance, with $t_{cap}$ the separation between the plates that encapsulate the dielectric, $A_{cap} = l_{cap}^2$ is the area of the cross-section of the smallest plate (see Fig. 3), and $\varepsilon_{cap}$ is the relative permittivity of the nonlinear dielectric. Note that one of the plates of the capacitor is coincident with the bottom patch of the two-sided mushroom structure.

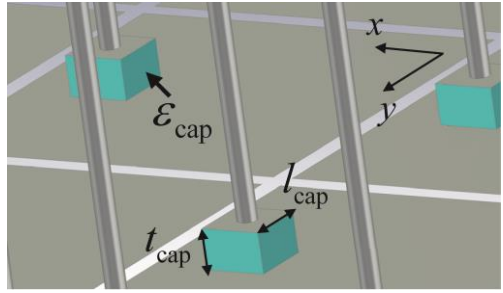

**Fig. 3** A nonlinear load connects the wires to the metallic patch at the bottom interface. The load is a parallel plate-type capacitor. The smallest plate has dimensions $l_{cap} \times l_{cap}$, and the plates are separated by a distance $t_{cap}$. The filling material is a Kerr-type nonlinear dielectric.

In the adopted framework, the nonlinear effects are treated perturbatively such that the frequency dependent impedance becomes a nonlinear function of the fields amplitude. The dominant field component in the capacitor region is evidently along the z-direction ($E_{z,L}$). Hence, it is assumed that $\varepsilon_{cap} = \varepsilon_{cap}^0 \left(1 + \alpha \left|E_{z,L}\right|^2\right)$ where $\alpha$ determines the strength of the nonlinear response and $\varepsilon_{cap}^0$ is the relative permittivity in the linear regime. The parameter $\alpha$ is proportional to the third order electric susceptibility $\chi^{(3)}$ of the material, and in a time-harmonic regime may be estimated as $\alpha = \dfrac{3}{4}\dfrac{\chi^{(3)}}{\varepsilon_{cap}^0}$ [54]. Within the validity of the model, the electric field inside capacitor is nearly constant and approximately equal to $E_{z,L} \approx V_L/t_{cap}$. Thus, the impedance of the load depends on the voltage drop at its terminals $Z_L = Z_L(V_L)$, and thereby also on the amplitude of the incident field.

Clearly, to have a strong nonlinear response the system should be desirably operated near some resonance so that $V_L$ (or equivalently the load current $I_L = I(z=-h)$) can be strongly dependent on small variations of the load impedance. To determine a suitable operating frequency, we used the analytical model of Appendix A to calculate $I_L$ as a function of frequency in the linear regime. In the analytical model, the load is modeled as a capacitor with $C = \varepsilon_0 \varepsilon_{cap}^0 A_{cap}/t_{cap}$. The structural parameters used in our simulations are $t_{cap} = 2g$, $A_{cap} = 9g^2$ and $\varepsilon_{cap}^0 = 2$, which yield $C = 0.04\text{pF}$. Figure 4 represents the load current as a function of frequency obtained using both the effective medium model [Eq. (A5)] and the full-wave electromagnetic simulator [57].

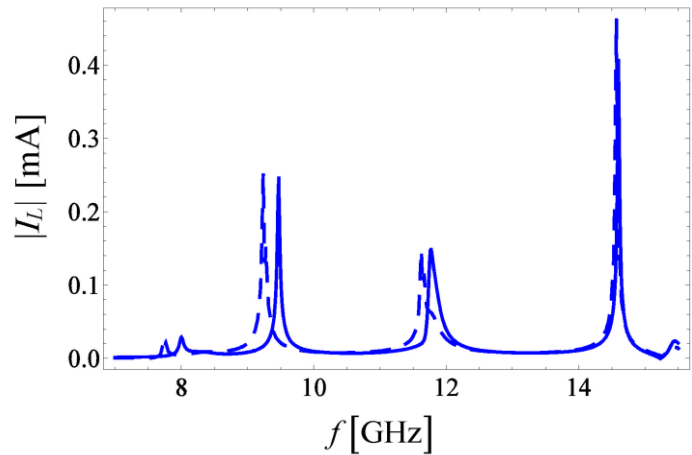

**Fig. 4** Amplitude of the electric current feeding the parallel-plate capacitor for an incident field with $E^{inc} = 1\text{V/m}$ (linear regime). Solid lines: homogenization model; Dashed lines: full-wave simulations [57].

Similar to Fig. 2, on the overall there is a reasonable agreement between the two calculation methods. Moreover, the results of Fig. 4 reveal that there are multiple resonances for which the current delivered to the lumped element is strongly enhanced. The strongest resonance occurs at 14.56GHz, i.e., the Fano resonance of Fig. 2b, and will be the focus of our attention in the following.

### III. Bistable Response

In the nonlinear regime, the relation between $V_L$ and the incident field is evidently nontrivial. Importantly, because the nonlinearity of the system is concentrated in the response of the lumped element, the nonlinear problem may be regarded as equivalent to a linear problem with an unknown impedance $Z_L$. This means that in the nonlinear regime the transfer function $V_L / E^{inc}$ depends only on the unknown value of the load $Z_L$, and thus can be rigorously determined based on the linear response of the system (even though $Z_L$ is a nonlinear function of $V_L$). We define the function $F^0(Z_L)$ as the ratio between the incident field and the voltage at a lumped load with a given impedance $Z_L$, calculated in the linear regime,

$$F^0(Z_L) = \left. \frac{E^{inc}}{V_L} \right|_{\text{linear regime}}. \tag{3}$$

For a fixed frequency, the function $F^0(Z_L)$ can be numerically evaluated using the homogenization theory of Appendix A. Then, in the nonlinear regime, the unknown $Z_L$ is found by numerically solving the nonlinear equation:

$$E^{inc} = F^0(Z_L) V_L, \quad \text{with } Z_L = Z_L(|V_L|). \tag{4}$$

The above formula makes clear that both $V_L$ and $Z_L$ are nonlinear functions of the incident field. The main requirement to obtain a bistable response is that in the linear regime the relation between the input ($|E^{inc}|$) and the output ($|V_L|$) depends on some parameter in a non-monotonic manner. From the relation $|V_L| = |F^0(Z_L)|^{-1} |E^{inc}|$, it follows that the transfer function $|F^0|$ must have a non-monotonic dependence on the load impedance. In these conditions, if the load impedance becomes a nonlinear function of the output, then the system response becomes bistable.

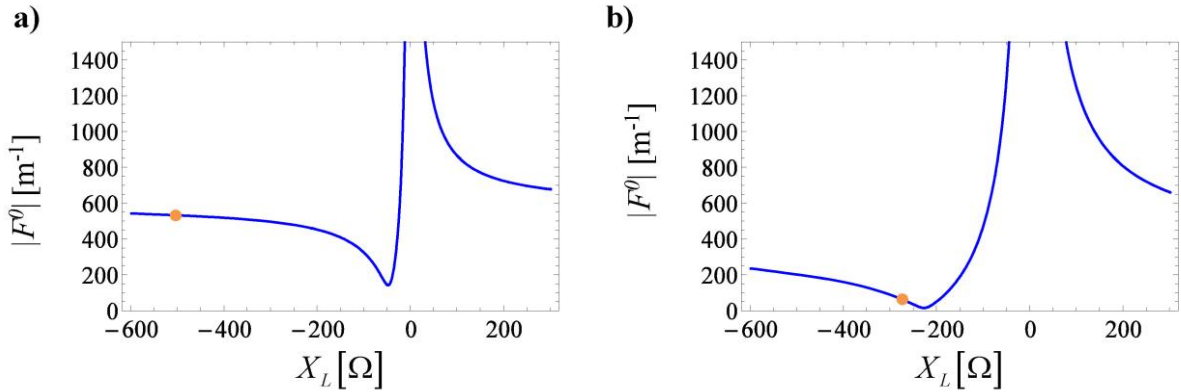

**Fig. 5** $|F^0|$ as a function of the reactance of the lumped element of the two-sided mushroom under the plane-wave excitation at a frequency: a) $f = 7.90 \text{GHz}$ and b) $f = 14.56 \text{GHz}$. The point of operation is marked with an orange dot.

To illustrate these ideas, we depict in Fig. 5 the numerically calculated transfer function $F^0$ as a function of the load reactance $X_L$ ($Z_L = -iX_L$) for the frequencies $f = 7.90 \text{GHz}$ (1$^{\text{st}}$ resonance in Fig. 4) and $f = 14.56 \text{GHz}$ (4$^{\text{th}}$ resonance in Fig. 4). The point of operation in the linear regime is marked with an orange circle in the plots. Evidently, the reactance of the lumped element has a strong influence on the transfer function. Importantly, for capacitive loads ($X_L < 0$), the monotonic behavior of $|F^0|$ changes in the vicinity of $X_L \approx -60\Omega$ and $X_L \approx -240\Omega$, for Figs. 5a and 5b, respectively. For $f = 14.56 \text{GHz}$ the point wherein $|F^0|$ has a minimum ($X_L \approx -i\,274\Omega$) is relatively close to the value of the load reactance in the linear regime. Hence, the Fano resonance at $f = 14.56 \text{GHz}$ enables us to operate the system near the point wherein $|F^0|$ changes the monotonic response, which is the basis of a bistable response as previously discussed. Thus, from here on we fix $f = 14.56 \text{GHz}$.

The nonlinear relation between $V_L$ and $Z_L$ can be found from Eq. (4). In particular, Eq. (4) implies that $|E^{inc}| = |F^0(Z_L(|V_L|))||V_L|$ and hence $|E^{inc}|$ can be regarded as a function of the load voltage, as shown in Fig. 6. In the simulation it was assumed that the nonlinear Kerr material is modeled by $\chi^{(3)} = 0.9 \times 10^{-9} \text{m}^2\text{V}^{-2}$. The results of Fig. 6 reveal that in a stationary regime the relation between the incident field and the load voltage is not univocal, and in particular for some values of $|E^{inc}|$ the load voltage $|V_L|$ has three distinct allowed values. To illustrate the consequences of such property, let us suppose that $|E^{inc}|$ is slowly increased from zero to a very large value. In that case, the load voltage is forced to suffer a discontinuous jump represented by the path $p_1$ in Fig. 6. On the other hand, if one slowly decreases $|E^{inc}|$ from very high values down to zero, a similar phenomenon occurs, corresponding to the path $p_2$ in Fig. 6. Therefore the point of operation of the system in a stationary regime follows a hysteresis-type loop, so that a given value of $|E^{inc}|$ can be associated with different voltages at the nonlinear load. Thus, the system may have a bistable response that depends on its past history. Similar effects have been reported in the literature for the refractive index of nonlinear metamaterials and plasmonic systems (see for example Refs. [48, 49]).

Furthermore, it is shown in Appendix B that much stronger nonlinearities may be implemented at microwave frequencies with commercially available varactors. Specifically, in Appendix B we compare the $E^{inc}$ vs. $V_L$ curves of a varactor and of the considered Kerr-material capacitor and show that the varactor has an increased sensitivity to the variations of the load voltage. It is demonstrated that the bistable response provided by a commercial varactor is approximately equivalent to that of a nonlinear parallel plate capacitor filled with a material with a $\chi^{(3)}$ parameter that may be up to three orders of magnitude larger than the value considered here. Note that our study is based on Kerr type nonlinearities to enable a direct full-wave validation of the results using CST Microwave Studio [57] (this commercial software does not support the nonlinear dynamics of a varactor).

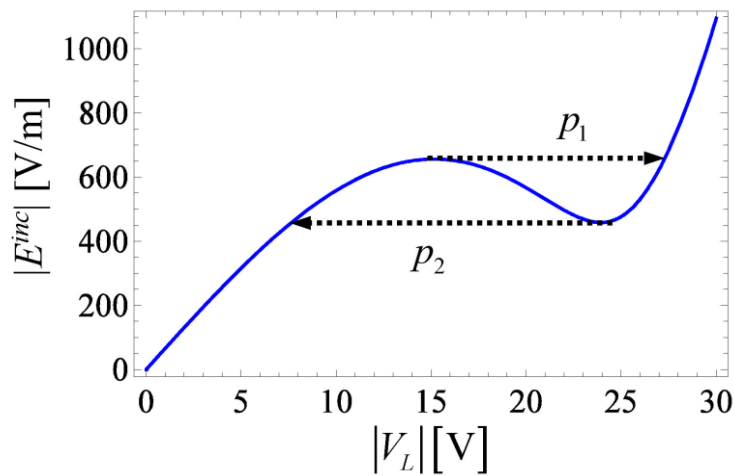

**Fig. 6** Incident electric field as a function of the voltage at the nonlinear load for the operating frequency 14.56 GHz.

Evidently, the hysteresis cycle of Fig. 6 implies that the scattering parameters $|R|$ and $|T|$ also follow analogous hysteresis loops as the amplitude of the incident field is varied. This is supported by Fig. 7, which depicts the hysteresis cycles calculated with the effective medium theory and with CST Microwave Studio [57]. The hysteresis loop is obtained with the full wave simulator by computing the scattering coefficients in the nonlinear regime with a dynamic excitation whose amplitude varies slowly in time. Specifically, the transmission and reflection properties are numerically determined by progressively increasing the excitation field from zero to any given amplitude, or by gradually decreasing the excitation field from high values to a desired value, depending on the branch of the hysteresis loop. As seen, there is a remarkable agreement between the effective medium theory and the full wave simulations.

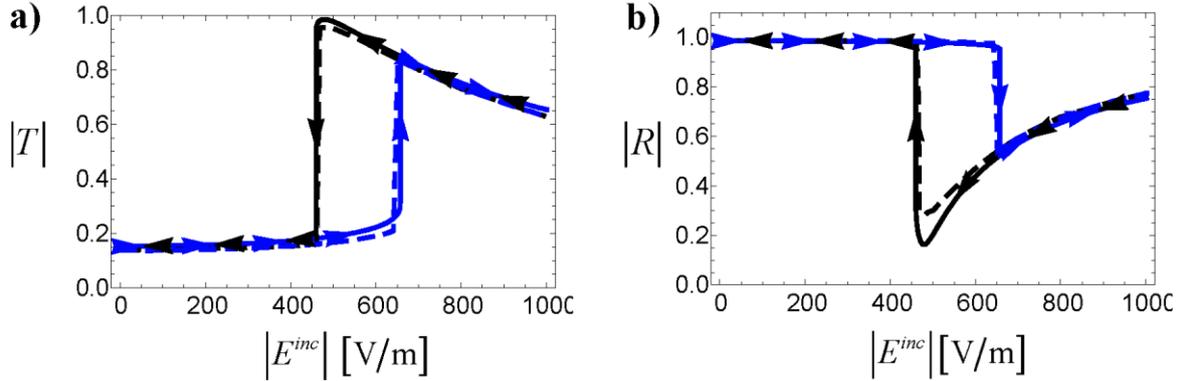

**Fig. 7** Profile of the **a)** transmission and **b)** reflection coefficients for the two-sided mushroom structure as function of the amplitude of the incident field. The black (blue) curves correspond to a scenario where the amplitude of the incident field decreases (increases) from a very high value to zero (from zero to a very high value). Solid lines: homogenization model. Dashed lines: full wave simulations [57]. The structural parameters and operating frequency are the same as in Fig. 6.

Most importantly, Fig. 7 reveals that the transitions between different branches are accompanied by a dramatic change in the values of the transmission and reflection coefficients, with a nonlinear modulation depth that can exceed 80%. In particular, the system can be switched between a state of nearly total transmission and another state where it is nearly opaque, depending on whether the incident field is increasing or decreasing. Typical from a bistable switch, our nonlinear metamaterial exhibits a jump-up characteristic, such that when the increasing (decreasing) incident field reaches some threshold, the transmission coefficient jumps up (down).

## IV. Dynamical Response

A bistable regime can have interesting practical applications in high-power microwave circuits. In our system the hysteresis loop of the scattering parameters is characterized by states with either a nearly total transmission or an almost complete reflection, which may enable the realization of an electromagnetic switch. To further explore this possibility, we performed a full wave simulation where the transmitted field was monitored as the amplitude of the incident field was dynamically varied.

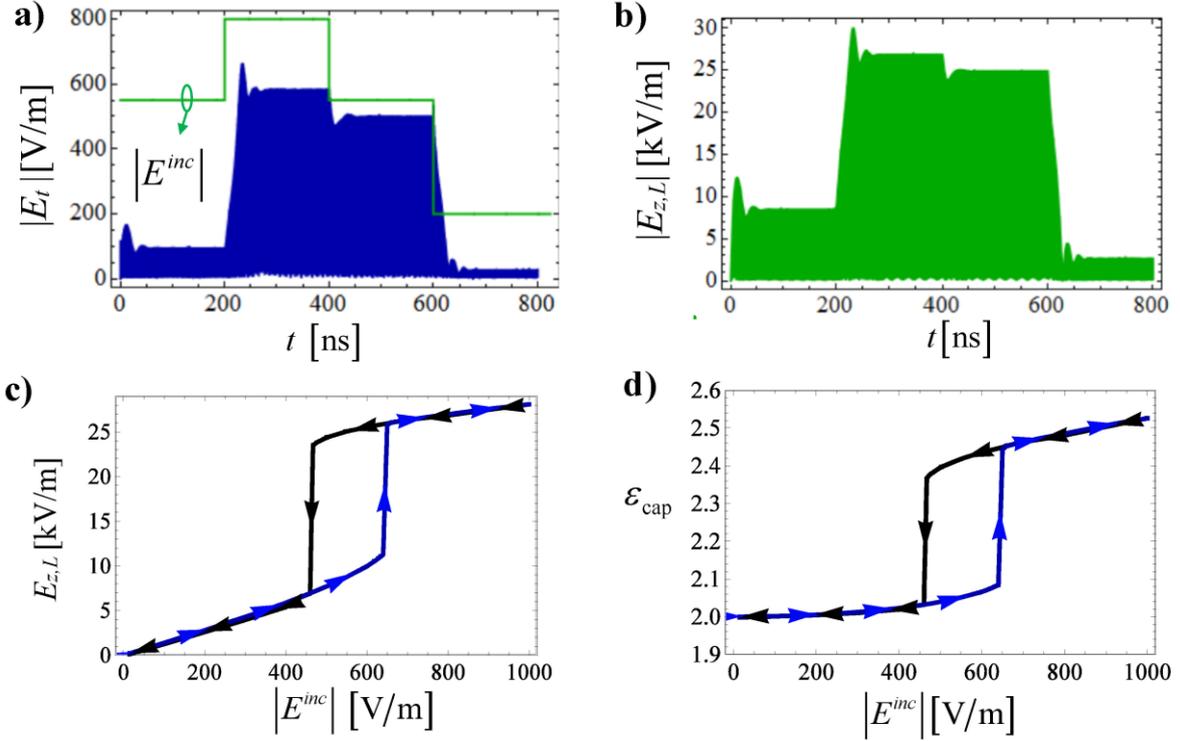

**Fig. 8 a)** Transmitted electric field (blue line) and as function of time when the incident field is a time varying modulated signal with carrier at $f = 14.56\text{GHz}$ and envelope (green line) such that $|E^{inc}| = 550\,\text{V/m}$ for $t < 200\text{ns}$, $|E^{inc}| = 800\,\text{V/m}$ for $200\text{ns} < t < 400\text{ns}$, $|E^{inc}| = 550\,\text{V/m}$ for $400\text{ns} < t < 600\text{ns}$, and $|E^{inc}| = 200\,\text{V/m}$ for $t > 600\text{ns}$. **b)** Similar to **a)** but for the z-component of the electric field inside the capacitor. **c)** and **d)** $|E_{z,L}|$ in the parallel-plate capacitor and the nonlinear permittivity $\varepsilon_{cap}$, respectively, as function of the incident field amplitude. The black curves correspond to a scenario where the amplitude of the incident field decreases from a very high value to zero, whereas the blue curves correspond to the reversed situation.

In the simulation, the incident wave is a time-varying modulated signal with an envelope (green line in Fig. 8a) that permits exploring the full hysteresis cycle of the nonlinear metamaterial. The transmitted field was calculated with Microwave Studio[2] and is represented with a blue line in Fig. 8a. The full wave simulation demonstrates unequivocally that the transmitted field depends on the past history of the system. Consistent with the hysteresis loop shown in Fig. 7, for an incident field with amplitude of $|E^{inc}| = 550$ V/m (1$^{st}$ and 3$^{rd}$ plateaus in Fig. 8a) the transmitted field depends on whether the incident field is increasing from very small values (the incoming wave is strongly reflected), or decreasing from high values (transmission coefficient is near unity). Furthermore, in the 2$^{nd}$ plateau, which corresponds to the strongest incident field, the transmission level (in percentage) is smaller than in the 3$^{rd}$ plateau. Note that the transmitted field oscillates very fast on the time scale of the plot (the period of the incident wave is 68.7 ps), and this is why it is represented as a shaded area.

The bistable response occurs due to the nonlinear dynamics of the system, allowing the structure to maintain a "memory" of the external excitation. This is further highlighted in Fig. 8b), which represents the time evolution of the field $E_{z,L}$ in the parallel plate capacitor, revealing dynamic features analogous to those of the transmitted field. Remarkably, the electromagnetic field inside the capacitor can be rather strong, and its amplitude can exceed 30 times that of the incident field. This property is the key to boost the nonlinear response and enhance the sensitivity of the nonlinear element to the incident field. Moreover, both $E_{z,L}$ and the permittivity of the Kerr material ($\varepsilon_{cap} = \varepsilon_{cap}^0 \left(1 + \alpha |E_{z,L}|^2\right)$) exhibit hysteresis loops with respect to the amplitude of the incident field (see Figs. 8c and 8d). The required nonlinear modulation of the permittivity is on the order of 25%. It was shown in early experiments of nonlinear metamaterials [38], that the rectification of the microwave signal may lead to charge accumulation at the terminals of a nonlinear lumped element, and thereby to a slowly time-varying biasing voltage. We numerically checked (not shown) with a two pulse repetition of the incident field that the response of our system appears to be weakly sensitive to such an effect. Yet, it is unclear if the numerical software can accurately capture the discussed phenomenon. In general, such a problem may be corrected with a large shunting inductor.

---

[2] Our $\chi^{(3)}$ parameter seems to be related to the definition used by CST Microwave Studio as: $\chi^{(3)} = 8/9 \chi_{CST}^{(3)}$.

## V. Conclusions

We developed an effective medium approach to study the response of asymmetric mushroom-type metamaterials loaded with nonlinear elements. It was shown that the dynamic variation of the impedance of the nonlinear element causes the metamaterial to exhibit a strong bistable response and hysteresis loops, and allows the mushroom structure to behave as an electromagnetic switch controlled by the intensity of the incoming field. In particular, a parallel-plate capacitor filled with a nonlinear Kerr-type material in the wires-to-patch connection enables the system to be switched between "go" and "no-go" transmission states, controlled by the dynamics of the incident field and dependent on the past history of the system. The operation near the Fano resonance boosts the nonlinear effects due to a strong enhancement of the electric field in the capacitor and due to the high sensitivity of the Fano resonance to changes in the capacitance. For the proposed design, the switch operation requires an incident field with an intensity on the order of $850 \,\text{W/m}^2$. As shown in Appendix B, our design is somewhat conservative and it appears that with a varactor it may be possible to implement much stronger nonlinearities with an "equivalent" $\chi^{(3)}$ increased by two or three orders of magnitude. In such a case, the electromagnetic switch operation may be attained with a power density on the order of $1 \,\text{W/m}^2$. Thus, our findings indicate that mushroom-type metamaterials can be interesting platforms to implement novel tunable nonlinear functionalities in the microwave regime.

## Appendix A: The effective medium model

The effective response of a set of infinitely long parallel metallic wires oriented along the $z$-direction and arranged in a periodic square lattice is described by the following dielectric function [1, 5]:

$$\overline{\overline{\varepsilon}}_{eff}(\omega, k_z) = \varepsilon_h \{\varepsilon_t (\hat{\mathbf{x}}\hat{\mathbf{x}} + \hat{\mathbf{y}}\hat{\mathbf{y}}) + \varepsilon_{zz}\hat{\mathbf{z}}\hat{\mathbf{z}}\}, \tag{A1}$$

where $\varepsilon_h$ is the dielectric host permittivity, $\varepsilon_{zz} = 1 + \left[\frac{\varepsilon_h}{(\varepsilon_m - \varepsilon_h)f_V} - \frac{k_h^2 - k_z^2}{k_p^2}\right]^{-1}$ and the transverse permittivity satisfies $\varepsilon_t \approx 1$ [1, 5]. Here, $k_h = \omega\sqrt{\varepsilon_h \mu_0}$ is the wave number in the host medium, $f_V = \pi(r_w/a)^2$ is the volume fraction of the metal in the cell, and $\varepsilon_m$ is the complex permittivity of the metallic wires. In this work, we suppose for simplicity that the metal is a PEC such that $\varepsilon_m = -\infty$ and $\varepsilon_{zz} = 1 - k_p^2/(k_h^2 - k_z^2)$. The parameter $k_p$ has the physical meaning of the plasma wave number of the effective medium, and within a thin wire approximation it can be written as $(k_p a)^2 = 2\pi \left[\ln\left(\frac{a^2}{4r_w(a-r_w)}\right)\right]^{-1}$ [15]. The explicit dependence of the dielectric function on the wave vector $k_z = -i\frac{\partial}{\partial z}$ implies a strong non-local behavior [1, 5].

Let us consider a transverse magnetic (TM) polarized incoming plane wave (magnetic field is along the $y$-direction) that impinges on the bilayer mushroom structure with an incidence angle $\theta_{inc}$ (see Fig. 1), so that the plane of incidence is the $xoz$ plane. As discussed in Ref. [29], a TM-polarized wave can excite plane waves in the wire medium slab with a transverse wave vector $\mathbf{k}_t = k_x\hat{\mathbf{x}} + k_y\hat{\mathbf{y}}$ such that $k_x = k_0 \sin\theta_{inc}$ and $k_y = 0$, with $k_0 = \omega/c$ and $c$ the light speed in the vacuum. In our problem, the electromagnetic field inside the wire medium slab can be written as a superposition of two pairs of counter-propagating waves (propagating along +z and –z directions, respectively) associated with the so-called transverse electromagnetic (TEM) and TM modes. The relevant field components in the scenario of interest are $H_y$, $E_x$ and $E_z$ and when the slab is surrounded by air, the electromagnetic fields satisfy [29]:

$$H_y = e^{ik_x x} \frac{E^{inc}}{\eta_0} \begin{cases} \left(e^{\gamma_0 z} - Re^{-\gamma_0 z}\right) & z > 0 \\ A_1 e^{\gamma_{TM}(z+h)} + A_2 e^{-\gamma_{TM}(z+h)} + B_1 e^{\gamma_{TEM}(z+h)} + B_2 e^{-\gamma_{TEM}(z+h)} & -h < z < 0 \\ Te^{\gamma_0(z+h)} & z < -h \end{cases} \quad (A2)$$

$$E_x = e^{ik_x x} \frac{E^{inc}}{\eta_0} \frac{1}{i\omega\varepsilon_0} \begin{cases} \gamma_0 \left(e^{\gamma_0 z} + Re^{-\gamma_0 z}\right) & z > 0 \\ \frac{\varepsilon_0}{\varepsilon_h} \gamma_{TM} \left(A_1 e^{\gamma_{TM}(z+h)} - A_2 e^{-\gamma_{TM}(z+h)}\right) + \frac{\varepsilon_0}{\varepsilon_h} \gamma_{TEM} \left(B_1 e^{\gamma_{TEM}(z+h)} - B_2 e^{-\gamma_{TEM}(z+h)}\right) & -h < z < 0 \\ \gamma_0 T e^{\gamma_0(z+h)} & z < -h \end{cases}$$

(A3)

$$E_z = -e^{ik_x x} \frac{E^{inc}}{\eta_0} \frac{k_x}{\omega\varepsilon_0} \begin{cases} e^{\gamma_0 z} - Re^{-\gamma_0 z} & z > 0 \\ \frac{\varepsilon_0}{\varepsilon_{zz}^{TM}} \left(A_1 e^{\gamma_{TM}(z+h)} + A_2 e^{-\gamma_{TM}(z+h)}\right) & -h < z < 0 \\ Te^{\gamma_0(z+h)} & z < -h \end{cases} \quad (A4)$$

where $R$ and $T$ are the reflection and transmission coefficients, and $A_{1,2}$ and $B_{1,2}$ represent the complex amplitudes of the TM and TEM waves, respectively, in the wire medium. The propagation constants of the TM and TEM modes are given by $\gamma_{TM} = \sqrt{k_x^2 + k_p^2 - \omega^2 \varepsilon_h \mu_0}$ and $\gamma_{TEM} = -i\omega\sqrt{\varepsilon_h \mu_0}$. The free space propagation constant along the z-direction is $\gamma_0 = \sqrt{k_x^2 - \omega^2 \mu_0 \varepsilon_0}$, $\eta_0$ is the free space impedance, and $E^{inc}$ is the incident field complex amplitude. The parameter $\varepsilon_{zz}^{TM} = \varepsilon_h k_x^2 / \left(k_p^2 + k_x^2\right)$ represents the z-component of the permittivity for the TM polarized wave. Moreover, the electric current $I$ that flows along the metallic wires can be written in terms of the macroscopic electromagnetic field as [16, 29]:

$$I = \left[ +i\omega\varepsilon_h E_z + \hat{\mathbf{z}} \cdot (i\mathbf{k}_t \times \mathbf{H}_t) \right] a^2 \tag{A5}$$

where $\mathbf{k}_t = k_x \hat{\mathbf{x}}$ and $\mathbf{H}_t = \mathbf{H} = H_y \hat{\mathbf{y}}$ represent the transverse (to the $z$-axis) components of the wave vector and magnetic field.

The unknowns $A_{1,2}$, $B_{1,2}$, $R$, and $T$ can be determined by imposing suitable boundary conditions at the interfaces. It was shown in [29] that when the wires are misaligned with respect to the patch grids, the pertinent boundary conditions at an interface $z = z_0$ with air are:

$$\lfloor E_x \rfloor_{z=z_0} = 0, \tag{A6a}$$

$$\lfloor H_y \rfloor_{z=z_0} = -Y_g E_x \big|_{z=z_0} + \frac{f_\alpha}{a} \hat{\mathbf{n}} \cdot \hat{\mathbf{z}} I, \tag{A6b}$$

$$\frac{\partial I}{\partial z} + \left( \frac{C_w}{C_{patch}} - i\omega C_w Z_{L,ef} \right) I (\hat{\mathbf{n}} \cdot \hat{\mathbf{z}}) - i\omega C_w f_\alpha a E_x = 0, \tag{A6c}$$

where $\lfloor F \rfloor_{z=z_0} = F_{z=z_0^+} - F_{z=z_0^-}$ stands for the field discontinuity of $F$ at the pertinent interface and $\hat{\mathbf{n}}$ is the outward unit vector directed towards the exterior of the wire medium. Here, $Z_{L,ef}$ represents the effective load impedance (i.e., including parasitic effects), and $Y_g$ is the grid admittance that relates the induced surface current with the tangential electric field. For an array of wires terminated with metallic patches separated by a gap $g$, the admittance $Y_g$ is given by $Y_g = -i(\varepsilon_h + \varepsilon_0)(\omega a/\pi) \log \left[ \csc(\pi g/2a) \right]$ [15]. Moreover, $L_w = \frac{\mu_0}{2\pi} \log \left( \frac{a^2}{4 r_w (a - r_w)} \right)$ is the per unit of length (p.u.l.) wire inductance, $C_w = \varepsilon_h \mu_0 L^{-1}$ is the p.u.l. wire capacitance and $C_{patch} = (\varepsilon_h + \varepsilon_0) \pi (a - g) \left[ \log(\sec(\pi g/2a)) \right]^{-1}$ is the patch capacitance [15]. The parameter $f_\alpha = \frac{d}{a - g}$ is an angular filling factor varying in the range of $-\frac{1}{2} < f_\alpha < \frac{1}{2}$, and depends on the shift $d$ (along the positive $x$-direction) between the wire and patch grids [29].

Using Eqs. (A2), (A3) and (A4), and imposing the boundary conditions (A6) at the two interfaces with air, one obtains a 6×6 linear system of equations. The numerical solution of this linear system determines the six unknowns $R$, $T$, $A_{1,2}$ and $B_{1,2}$. The current at the load position ($I_L = I(z=-h)$) is determined using Eq. (A5), and the corresponding voltage drop is given by $V_L = Z_{L,ef} I_L$.

## Appendix B: Mushroom metamaterial loaded with varactors

As discussed in the main text, the emergence of a bistable regime requires a non-monotonic variation of $|F^0|$ with the impedance of the load. Next, it is shown that "varactors" can be an interesting solution to implement a hysteresis-type (bistable) response in mushroom-type metamaterials.

Varactors are nonlinear elements ubiquitous in microwave devices. The capacitance of a varactor depends on the voltage $V_L$ at its terminals. A typical model for the nonlinear capacitance is [62]:

$$C(V_L) = C_{J0}\left(1 + \frac{V_L}{V_J}\right)^{-M} + C_p, \tag{B1}$$

where $C_{J0}$ is zero-bias junction capacitance, $V_J$ is the junction potential, $M$ is the grading coefficient and $C_p$ is the package capacitance. Typical values for these parameters are $C_p = 50\,\text{fF}$, $C_{J0} = 2.92\,\text{pF}$, $V_J = 0.68\,\text{V}$ and $M=0.41$ [62]. These parameters will be used throughout this Appendix. In addition, a parasitic inductance $L_S = 1.9\,\text{nH}$ in series with the junction capacitance is also considered in the model of the varactor [62]. It should be noted that a varactor is a diode and thus conducts current in one of the directions. Hence, it practice one may need to use two oppositely directed varactors in series or alternatively heterostructure barrier varactors [63].

Using the same methods as in Sec. III, we determined the nonlinear relation between the incident field and the voltage $V_L$ when the varactor is used as the nonlinear element in the mushroom metamaterial. The calculated characteristic curve $E^{inc}$ vs. $V_L$ is represented in Fig. 9a and has features completely analogous to those obtained for a Kerr-type nonlinearity.

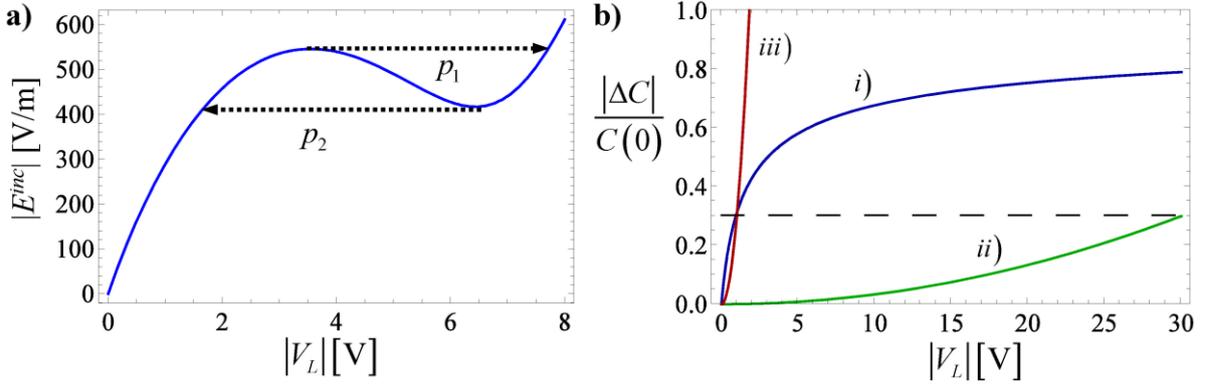

**Fig. 9 a)** Incident electric field as a function of the voltage at the nonlinear load for a varactor. The operation frequency is at $f = 7.20\text{GHz}$. **b)** Relative variation of the capacitance as a function of the voltage for: *i)* the varactor and *ii)* the nonlinear parallel-plate capacitor considered in the main text *iii)* a parallel-plate capacitor with the same structural parameters as in *ii)* but filled with a Kerr-type material with $\chi^{(3)} = 0.9 \times 10^{-6} \text{m}^2\text{V}^{-2}$.

Note that for a varactor the capacitance *decreases* with $|V_L|$, whereas for the Kerr-material capacitor considered in the main text the capacitance *increases* with $|V_L|$. The important point is that in both cases the capacitance varies monotonically with $|V_L|$. This property and the non-monotonic behavior of $|F^0|$ guarantee a bistable response provided the point of operation is carefully chosen. Indeed, similar to Fig. 6, the characteristic $E^{inc}$ vs. $V_L$ in Fig. 9a is multi-valued, so that the physical response of the varactor is forced to have discontinuous jumps corresponding to the paths $p_1$ and $p_2$. Thus, the bistable response can be obtained with different types of nonlinear elements. The operation frequency in Fig. 9a is lower than in Fig. 6 because the capacitance $C(0)$ in the linear regime is a few times larger for the varactor.

Notably, the considered varactor enables a nonlinear response much stronger than the nonlinear element of the main text. This is shown by Fig. 9b*i-ii*, which represents the relative change in the capacitance $|C(V_L) - C(0)|/C(0) \equiv |\Delta C|/C(0)$ as a function of the voltage at each nonlinear element. Clearly, the varactor is more sensitive to the variations of the voltage, particularly for low values of $|V_L|$. The hysteresis loops shown in Fig. 6 (which corresponds to $|V_L| < 30\text{V}$) require a nonlinear modulation of the capacitance on the order of 30% (marked as a dashed horizontal line in Fig. 9b). The same relative variation of the varactor capacitance can be achieved with a voltage as small as $|V_L| \approx 0.9\text{V}$. In order to mimic the response of the varactor for $|\Delta C|/C < 0.3$ one would need to use a parallel-plate capacitor with a $\chi^{(3)}$ parameter as large as $0.9 \times 10^{-6} \text{m}^2\text{V}^{-2}$ (Fig. 9biii).

**Acknowledgements:**
This work was funded by Fundação para a Ciência e a Tecnologia under project PTDC/EEI-TEL/4543/2014. D. E. Fernandes acknowledges support by Fundação para a Ciência e a Tecnologia, Programa Operacional Capital Humano/POCH, and the cofinancing of Fundo Social Europeu and MCTES under the fellowship SFRH/BPD/116525/2016.

# References:

[1] P. A. Belov, R. Marques, S. I. Maslovski, I. S. Nefedov, M. Silveirinha, C. R. Simovski, and S. A. Tretyakov, "Strong spatial dispersion in wire media in the very large wavelength limit", *Phys. Rev. B* **67**, 113103 (2003).
[2] S. Maslovski, S. Tretyakov, and P. Belov, "Wire media with negative effective permittivity: A quasi-static model", *Microwave Opt. Technol. Lett.* **35**, 47 (2002).
[3] S. A. Tretyakov, and S. I. Maslovski, "Thin absorbing structure for all incidence angles based on the use of a high-impedance surface", *Microwave Opt. Tech. Lett.* **38**, 175 (2003).
[4] M. G. Silveirinha, and C. A. Fernandes, "Homogenization of 3D- Connected and Non-Connected Wire Metamaterials", *IEEE Trans. on Microwave Theory and Tech.* **53**, 1418 (2005).

[5] M. G. Silveirinha, "Nonlocal homogenization model for a periodic array of ϵ-negative rods", *Phys. Rev. E* **73**, 046612 (2006).

[6] P. A. Belov, Y. Hao, and S. Sudhakaran, "Subwavelength microwave imaging using an array of parallel conducting wires as a lens", *Phys. Rev. B* **73**, 033108 (2006).

[7] M. G. Silveirinha, P. A. Belov, and C. Simovski, "Subwavelength Imaging at Infrared Frequencies Using an Array of Metallic Nanorods", *Phys. Rev. B* **75**, 035108 (2007).

[8] G. Shvets, S. Trendafilov, J. B. Pendry, and A. Sarychev, "Guiding, Focusing, and Sensing on the Subwavelength Scale using Metallic Wire Arrays", *Phys. Rev. Lett.* **99**, 053903 (2007).

[9] P. Ikonen, C. Simovski, S. Tretyakov, P. Belov, and Y. Hao, "Magnification of subwavelength field distributions at microwave frequencies using a wire medium slab operating in the canalization regime", *Appl. Phys. Lett.* **91**, 104102 (2007).

[10] A. Demetriadou, and J. B. Pendry, "Taming spatial dispersion in wire metamaterial", *Phys. Condens. Matter* **20**, 295222 (2008).

[11] J. Yao, Z. Liu, Y. Liu, Y. Wang, C. Sun, G. Bartal, A. M. Stacy, and X. Zhang, "Optical Negative Refraction in Bulk Metamaterials of Nanowires", *Science* **321**, 930 (2008).

[12] P. A. Belov, Y. Zhao, S. Tse, P. Ikonen, M. G. Silveirinha, C. R. Simovski, S. Tretyakov, Y. Hao, and C. Parini, "Transmission of images with subwavelength resolution to distances of several wavelengths in the microwave range", *Phys. Rev. B* **77**, 193108 (2008).

[13] O. Luukkonen, M. G. Silveirinha, A. B. Yakovlev, C. R. Simovski, I. S. Nefedov, and S. A. Tretyakov, "Effects of spatial dispersion on reflection from mushroom-type artificial impedance surfaces", *IEEE Trans. Microwave Theory Tech.* **57**, 2692 (2009).

[14] A. B. Yakovlev, M. G. Silveirinha, O. Luukkonen, C. R. Simovski, I. S. Nefedov, and S. A. Tretyakov, "Characterization of surface-wave and leaky-wave propagation on wire-medium slabs and mushroom structures based on local and nonlocal homogenization models", *IEEE Trans. Microwave Theory Tech.* **57**, 2700 (2009).

[15] S. I. Maslovski, and M. G. Silveirinha, "Nonlocal permittivity from a quasistatic model for a class of wire media", *Phys. Rev. B* **80**, 245101 (2009).

[16] S. I. Maslovski, T. A. Morgado, M. G. Silveirinha, C. S. R. Kaipa, and A. B. Yakovlev, "Generalized additional boundary conditions for wire media", *New J. Phys.* **12**, 113047 (2010).

[17] P. A. Belov, G. K. Palikaras, Y. Zhao, A. Rahman, C. R. Simovski, Y. Hao, and C. Parini, "Experimental demonstration of multiwire endoscopes capable of manipulating near-fields with subwavelength resolution", *Appl. Phys. Lett.* **97**, 191905 (2010).

[18] T. A. Morgado, J. S. Marcos, M. G. Silveirinha, and S. I. Maslovski, "Experimental Verification of Full Reconstruction of the Near-Field with a Metamaterial Lens", *Appl. Phys. Lett.* **97**, 144102 (2010).

[19] S. Paulotto, P. Baccarelli, P. Burghignoli, G. Lovat, G.W. Hanson, and A.B. Yakovlev, "Homogenized Green's functions for an aperiodic line source over planar densely periodic artificial impedance surfaces", *IEEE Trans. Microwave Theory Tech*. **58**, 1807 (2010).

[20] J. T. Costa, and M. G. Silveirinha, "Achromatic Lens Based on a Nanowire Material with Anomalous Dispersion", *Opt. Express* **20**, 13915 (2012).

[21] C. R. Simovski, P. A. Belov, A. V. Atrashchenko, and Y. S. Kivshar, "Wire metamaterials: physics and applications", *Adv. Mater.* **24**, 4229 (2012).

[22] D. E. Fernandes, S. I. Maslovski, and M. G. Silveirinha, "Cherenkov emission in a nanowire material", *Phys. Rev. B* **85**, 155107 (2012).

[23] V. V. Vorobev, and A. V. Tyukhtin, "Nondivergent Cherenkov radiation in a wire metamaterial", *Phys. Rev. Lett.* **108**, 184801 (2012).

[24] G. W. Hanson, E. Forati, and M. G. Silveirinha, "Modeling of Spatially-Dispersive Wire Media: Transport Representation, Comparison With Natural Materials, and Additional Boundary Conditions", *IEEE Trans. Antennas Propag.* **60**, 4219 (2012).

[25] B. M. Wells, A. V. Zayats, and V. A. Podolskiy, "Nonlocal optics of plasmonic nanowire metamaterials", *Phys. Rev. B* **89**, 035111 (2014).

[26] C. S. R. Kaipa, A. B. Yakovlev, S. I. Maslovski, and M. G. Silveirinha, "Indefinite dielectric response and all-angle negative refraction in a structure with deeply-subwavelength inclusions", *Phys. Review B* **84**, 165135 (2011).

[27] C. S. R. Kaipa, A. B. Yakovlev, S. I. Maslovski, and M. G. Silveirinha, "Mushroom-type High-Impedance Surface with Loaded Vias: Homogenization Model and Ultra-Thin Design", *IEEE Antennas and Wireless Propag. Lett.* **10**, 1503 (2011).

[28] F. Yang, and Y. Rahmat-Samii, "Polarization-dependent electromagnetic band gap (PDEBG) structures: Designs and applications", *Microwave and Opt. Tech. Lett.* **41**, 6 (2004).

[29] D. E. Fernandes, S. I. Maslovski, and M. G. Silveirinha, "Asymmetric Mushroom-Type Metamaterials", *IEEE Trans. Microwave Theory Tech.* **62**, 8 (2014).

[30] Y. Liu, G. Bartal, D. A. Genov, and X. Zhang, "Subwavelength Discrete Solitons in Nonlinear Metamaterials", *Phys. Rev. Lett.* **99**, 153901 (2007).

[31] F. Ye, D. Mihalache, B. Hu, and N. C. Panoiu, "Subwavelength Plasmonic Lattice Solitons in Arrays of Metallic Nanowires", *Phys. Rev. Lett.* **104**, 106802 (2010).

[32] M. G. Silveirinha, "Theory of Spatial Optical Solitons in Metallic Nanowire Materials", *Phys. Rev. B* **87**, 235115 (2013).

[33] D. E. Fernandes, and M. G. Silveirinha, "Bright and Dark Spatial Solitons in Metallic Nanowire Arrays", *Photon. Nanostruct.: Fundam. Appl.* **12**, 340 (2014).

[34] H. Wakatsuchi, S. Kim, J. Rushton, and D. Sievenpiper, "Circuit-based nonlinear metasurface absorbers for high power surface currents", *Appl. Phys. Lett.* **102**, 214103 (2013).

[35] Z. Luo, X. Chen, R. Quarfoth, and D. Sievenpiper, "Nonlinear Power Dependent Impedance Surface", *IEEE Trans. on Antennas and Propag.* **63**, 1736 (2015).

[36] S. Kim, H. Wakatsuchi, J. Rushton, and D. Sievenpiper, "Switchable Nonlinear Metasurfaces for Absorbing High Power Surface Waves", *Appl. Phys. Lett.* **108**, 041903 (2016).

[37] M. Lapine, M. Gorkunov, and K. H. Ringhofer, "Nonlinearity of a metamaterial arising from diode insertions into resonant conductive elements", *Phys. Rev. E* **93**, 065601(R) (2003).

[38] D. A. Powell, I. V. Shadrivov, Y. S. Kivshar, and M. V. Gorkunov, "Self-tuning mechanisms of nonlinear split-ring resonators", *Appl. Phys. Lett.* **91**, 144107 (2007).

[39] I. V. Shadrivov, A. B. Kozyrev, D. W. van der Weide, and Y. S. Kivshar, "Nonlinear magnetic metamaterials", *Opt. Express* **16**, 20266 (2008).

[40] D. A. Powell, I. V. Shadrivov, and Y. S. Kivshar, "Nonlinear electric metamaterials", *Appl. Phys. Lett.* **95**, 084102 (2009).

[41] B. Wang, J. Zhou, T. Koshny, and C. M. Soukoulis, "Nonlinear properties of split-ring resonators", *Opt. Express* **16**, 16058 (2008).

[42] R. Yang, and I. V. Shadrivov, "Double-nonlinear metamaterials", *Appl. Phys. Lett.* **97**, 231114 (2010).

[43] T. Nakanishi, Y. Tamayama, and M. Kitano, "Efficient second harmonic generation in a metamaterial with two resonant modes coupled through two varactor diodes", *Appl. Phys. Lett.* **100**, 044103 (2012).

[44] K. E. Hannam, D. A. Powell, I. V. Shadrivov, and Y. S. Kivshar, "Tuning the nonlinear response of coupled split-ring resonators", *Appl. Phys. Lett.* **100**, 081111 (2012).

[45] M. Lapine, I. V. Shadrivov, and Y. S. Kivshar, "Wide-band negative permeability of nonlinear metamaterials", *Sci. Rep.* **2**, 412 (2012).

[46] Y. Ding, C. Xue, Y. Sun, H. Jiang, Y. Li, H. Li, and H. Chen, "Subwavelength electromagnetic switch: Bistable wave transmission of side-coupling nonlinear meta-atom", *Opt. Express* **20**, 24813 (2012).

[47] M. Liu, Y. Sun, D. A. Powell, I. V. Shadrivov, M. Lapine, R. C. McPhedran, and Y. S. Kivshar, "Nonlinear response via intrinsic rotation in metamaterials", *Phys. Rev. B* **87**, 235126 (2013).

[48] P.-Y. Chen, M. Farhat, and A. Alù, "Bistable and Self-Tunable Negative-Index Metamaterial at Optical Frequencies", *Phys. Rev. Lett.* **106**, 105503 (2011).

[49] C. Argyropoulos, P.-Y. Chen, G. D'Aguanno, N. Engheta, and A. Alù, "Boosting optical nonlinearities in ε-near-zero plasmonic channels", *Phys. Rev. B* **85**, 045129 (2012).

[50] P.-Y. Chen, and A. Alù, "Optical nanoantenna arrays loaded with nonlinear materials", *Phys. Rev. B* **82**, 235405 (2010).

[51] I. Maksymov, A. Miroshnichenko, and Y. S. Kivshar, "Actively tunable bistable optical Yagi-Uda nanoantenna", *Opt. Express* **20**, 8929 (2012).

[52] H. Harutyunyan, G. Volpe, R. Quidant, and L. Novotny, "Enhancing the Nonlinear Optical Response Using Multifrequency Gold-Nanowire Antennas", *Phys. Rev. Lett.* **108**, 217403 (2012).

[53] A. M. Mahmoud, A. R. Davoyan, and N. Engheta, "All-passive nonreciprocal metastructure", *Nature Comms.* **6**, 8359 (2015).

[54] S. Lannebère, and M. G. Silveirinha, "Optical meta-atom for localization of light with quantized energy", *Nature Comms.* **6**, 8766 (2015).

[55] M. Lapine, I. V. Shadrivov, and Y. S. Kivshar, "Colloquium: Nonlinear metamaterials", *Rev. Mod. Phys.* **86**, 1093 (2014).

[56] M. Lapine, "New degrees of freedom in nonlinear metamaterials", *Phys. Status Solidi B* **254**, 1600462 (2017).

[57] CST GmbH 2017 CST Microwave Studio http://www.cst.com.

[58] U. Fano, "Effects of Configuration Interaction on Intensities and Phase Shifts", *Phys. Rev.* **124**, 1866 (1961).

[59] A. E. Miroshnickenko, S. Flach, and Y. S. Kivshar, "Fano resonances in nanoscale structures", *Rev. Mod. Phys.* **82**, 2257 (2010).

[60] B. Luk'yanchuk, N. I. Zheludev, S. A. Maier, N. J. Halas, P. Nordlander, H. Giessen, and C. T. Chong, "The Fano resonance in plasmonic nanostructures and metamaterials", *Nat. Mater.* **9**, 707 (2010).

[61] D. E. Fernandes, S. I. Maslovski, and M. G. Silveirinha, "Fano resonances in nested wire media", *Phys. Rev. B* **88**, 045130 (2013).

[62] Skyworks, *Varactor SPICE Models for RF VCO Applications*, SMV APN1004 datasheet (2005).

[63] J. Carbonell, V. E. Boria, and D. Lippens, "Nonlinear Effects in Split Ring Resonators Loaded with Heterostructure Barrier Varactors", *Microw. Opt. Technol. Lett*. **50**, 474 (2008).